\documentclass[12pt]{iopart}
\newcommand{\gguide}{{\it Preparing graphics for IOP journals}}
\usepackage{epsfig}
\usepackage{graphicx}
\usepackage{bm}
\def\br{\begin{eqnarray}}
\def\er{\end{eqnarray}}
\def\be{\begin{equation}}
\def\ee{\end{equation}}

\def\({\left(}
\def\){\right)}

\def\<{\left\langle}
\def\>{\right\rangle}

\begin{document}

\title[How confinement may affect technicolor?]{How confinement may affect technicolor?}

\author{A.~Doff $^1$,  F. A. Machado $^2$ and A. A. Natale ${}^2$}

\address{$^1$ Universidade Tecnol\'ogica Federal do Paran\'a - UTFPR - COMAT \\
Via do Conhecimento Km 01, 85503-390, Pato Branco - PR, Brazil}
\address{$^2$ Instituto de F\'{\i}sica Te\'orica, UNESP - Universidade Estadual Paulista, \\
Rua Dr. Bento T. Ferraz, 271, Bloco II,
01140-070, S\~ao Paulo - SP,
Brazil}
\ead{agomes@utfpr.edu.br}

\begin{abstract}
Confinement has been introduced into the quark gap equation, as proposed by Cornwall, as a possible solution to the problem
of chiral symmetry breaking in QCD with dynamically massive gluons. We argue that the same mechanism can be applied for technicolor
with dynamically massive technigluons. Within this approach both theories develop a hard self-energy dynamics,
resulting from an effective four-fermion interaction, which does not lead to the known technicolor phenomenological problems. 
We outline a quite general type of 
technicolor model within this proposal that may naturally explain the masses of different fermion generations.
\end{abstract}

\maketitle

\section{Introduction}

\par The nature of the electroweak symmetry breaking is one of the most important problems in
particle physics, and there are many questions that may be answered in the near future by the LHC
experiments, such as the following: Is the Higgs boson, if it exists at all, elementary or composite,
and what are the symmetries behind the Higgs mechanism? The possibility that the Higgs boson is a composite
state instead of an elementary one is more akin to the phenomenon of spontaneous symmetry breaking that
originated from the Ginzburg-Landau Lagrangian, which can be derived from the microscopic BCS theory of
superconductivity describing the electron-hole interaction (or the composite state in our case). This
dynamical origin of the spontaneous symmetry breaking has been discussed with the use of many models,
the technicolor (TC) being the most popular one \cite{sannino}.

Ordinary fermion masses 
($m_f$) result from the interaction of these fermions with technifermions through the exchange of a extended technicolor boson (ETC) and depend crucially on the technifermion self-energy. In the early models this self-energy was considered to be given by the standard operator product expansion (OPE) 
result \cite{politzer}:
$\Sigma_{TC} (p^2) \propto \left\langle {\bar{T}_f}T_f\right\rangle/p^2$,
where $\left\langle {\bar{T}_f}T_f\right\rangle$ is the TC condensate and of order of a few hundred GeV. With this self-energy
the fermion masses are given by $m_f \approx \left\langle {\bar{T}_f}T_f\right\rangle/M_{etc}^2$. In order to obtain the
fermion masses of the second and third generations the ETC gauge boson masses had to be very light. Since these bosons connect
different fermionic generations and must be light, they may produce flavor changing neutral currents (FCNC) incompatible with the experimental data. A possible way out of this dilemma was proposed by Holdom \cite{holdom}, remembering that the self-energy behaves as
\be 
\Sigma_{TC} (p^2)\approx \frac{\left\langle {\bar{T}_f}T_f\right\rangle_\mu}{p^2} \left(\frac{-p^2}{\mu^2}\right)^{\gamma_m /2} \,\, ,
\ee
where $\mu$ is the characteristic TC scale and $\gamma_m$ the anomalous dimension associated to the condensate operator.

If $\gamma_m \geq 1$ the fermion masses have a smaller dependence on the ETC gauge boson masses, which can be larger resolving the FCNC problem. Models proposing such large
anomalous dimensions were reviewed in Ref.\cite{yama}. Therefore, theories with large anomalous dimensions ($\gamma_m$) are quite desirable for technicolor phenomenology \cite{sannino}. Lattice simulations were used to study many models that could have a
large $\gamma_m$, regrettably some of these models do have an appreciable anomalous dimension but not large enough to solve the phenomenological problems of TC theories \cite{note1}, indicating how difficult is to build a realistic TC model.

It is quite possible that the TC problems are related to the poorly known self-energy expression, or the way chiral
symmetry breaking (CSB) is realized in non-Abelian gauge theories.
Actually, the only known laboratory to test the CSB mechanism is QCD, and even in this case, considering several
recent results about dynamical mass generation in QCD that we shall discuss throughout the paper, imply that the dynamical quark mass generation mechanism is not fully understood. We will argue that a recent proposal to understand the CSB mechanism in
QCD \cite{cornwall3,cornwall4} may shed light on the same mechanism in TC models and possibly lead to viable theories for the
dynamical symmetry breaking of the standard model.

The effect of confinement 
in TC, as well as in QCD, may be so strong that an effective four-fermion interaction (like the famous Nambu--
Jona-Lasinio gauged model) can be generated. The theory develops a quite hard self-energy implying that FCNCs can be avoided by decoupling the techni-gauge fields, while different family fermion masses are generated via chiral symmetry breaking of the TC and QCD theories.

In Section II of this work we will recall that QCD possess the property of dynamical generation of gluon masses. This property
has been verified through lattice simulations as well as through Schwinger-Dyson calculations. We argue that the same mechanism
happens in TC theories (i.e. generation of dynamically massive technigluons), as long as the TC model is not conformal. If this is the case the same CSB problem appearing in QCD will appear in TC: QCD (TC) with dynamically massive (techni)gluons does not have
strength enough to generate the expected (techni)quark masses, or (techni)quark condensates. A possible solution for this problem,
as proposed by Cornwall \cite{cornwall3,cornwall4} and followed by us \cite{we}, is that confinement is responsible for the CSB in QCD,
and we assume this to be also true for TC. In Section III we discuss how we can model confinement in non-Abelian theories in order
to obtain the right amount of chiral symmetry breaking. We argue that in QCD, as well in TC, the confinement effect is
so strong that can generate an effective four-fermion interaction (or an effective Nambu-Jona-Lasinio gauged model). This
result in quite hard self-energies for quarks as well as techniquarks. In Section IV we discuss how QCD and TC theories
with hard self-energies, or self-energies that have been called in the past as irregular solutions for $\Sigma (p^2)$, may
lead to models where both theories contribute to the ordinary fermion masses, and do not lead to FCNC problems. Section
V contains our conclusions. 

\section{Technicolor with dynamically massive technigluons}

\par Many years ago Cornwall proposed that a dynamical gluon mass could be generated in QCD \cite{cornwall}. Only recently this possibility
was confirmed by lattice simulations \cite{cucchi} and checked rigorously through Schwinger-Dyson equations (SDE) \cite{aguilar}. It seems 
that this is a general property of non-Abelian gauge theories \cite{cornwalll}. There is no reason to believe that the same mechanism
does not happen in TC theories. The only possibility for technigluons not acquiring a dynamical mass that we can think of is the case of 
a conformal or non-asymptotically free TC model, where the effect of technifermion loops in the Schwinger-Dyson equations cancel the gauge
loop effects responsible for the dynamical technigluon mass.

We assume a TC theory based, for instance, on a $SU(N)$ gauge group, with a fermion content such that the theory is asymptotically free and
is not almost conformal (or not near a perturbative fixed point). We also assume that in this theory the technigluons will acquire a dynamical mass, and CSB breaking
can be studied in the same way it is studied for QCD, through the Schwinger-Dyson equations for the technifermions. The techifermion
self-energy will be given by:
\br
&&\Sigma_{TC}\equiv M(p^2) = \nonumber \\
&&\frac{C_2 }{(2\pi)^4}\int  d^4k   \frac{{\bar{g}_{tc}}^2(p-k)3M(k^2)}{[(p-k)^2+m_{tg}^2(p-k)][k^2+M^2(k^2)]},
\label{eqn1}
\er
where we consider the Landau gauge, $C_2$ is the Casimir operator for the fermionic representation, $m_{tg}(k^2)$ is the dynamical technigluon mass and ${\bar{g}}_{tc}$ the effective TC coupling constant. First, we must say that, as far as we know, the CSB mechanism in TC models has not
been studied up to now in the presence of dynamical technigluon mass generation. Secondly, to understand what may happen in a TC theory
we will recall some QCD results. 

When dynamical gauge boson masses are generated in any asymptotically free non-Abelian gauge theory we also expect that the coupling
constant develops a non-perturbative infrared fixed point \cite{natale2a}. In QCD it was predicted many years ago that the coupling
constant would behave as \cite{cornwall} 
\be
{\bar{g}}^2_{QCD}(k^2)= \frac{1}{b \ln[(k^2+4m_g^2)/\Lambda_{QCD}^2]} \, ,
\label{eq04}
\ee
where $b$ is the first $\beta$ function coefficient, and $m_g\equiv m_g(k^2=0)\approx 2\Lambda_{QCD} \approx 500-600$ MeV (the phenomenologically preferred infrared value of the gluon mass \cite{natale}). This charge's value at the infrared fixed point ($\alpha_s (0)\equiv {\bar{g}}^2(0)/4\pi$) is of order $0.5$. This number may be considered surprisingly small but there are several phenomenological calculations indicating that this
value should not be larger than $1$; see, for instance, a compilation of infrared values of the QCD coupling constant shown in Ref.\cite{freez}.
Now, the gluon propagator in the fermionic SDE kernel, no longer behaves as $1/k^2$ but as $1/(k^2+m_g^2)$ in the infrared, what diminishes the
strength of the interaction, and we also add to this fact the damping caused by the small value of the infrared coupling constant (${\bar{g}}^2(0)$).
The consequence is that we do not generate dynamical quark masses ($M(k^2)$) (or quark condensates) compatible with the experimental data in QCD for quarks in the fundamental representation \cite{haeri}! QCD could only generate CSB if quarks were in higher dimensional representations, i.e. with higher values for the Casimir operator in order to compensate the infrared damping discussed above \cite{cornwall2}.

TC theories will also present dynamical technigluon mass generation, and for the same reasons that we discussed in the QCD case, i. e. a small
infrared TC coupling constant and the damping caused by the $1/m_{tg}^2$ infrared value of the technigluon propagator, we do not
expect that they will develop enough chiral symmetry breaking to form the TC condensates. 
In this work we will follow the idea of
Ref.\cite{cornwall3,cornwall4} that confinement is necessary and sufficient to promote CSB and develop the expected (techni)quark condensates.
Actually, our next section will start discussing evidences for a relation between CSB and confinement. 

\section{Chiral symmetry breaking as a consequence of confinement}

\par The majority of studies about CSB in gauge theories, no matter if in QCD or TC, relied on the one-gauge boson exchange. If we deal with
dynamically massive gauge bosons, as discussed in the previous section, CSB will not be achieved at least if we have fermions in the
fundamental representation. We will than consider the case where confinement is necessary for CSB, and in order to emphasize this 
possibility we will review some QCD aspects that point out in this direction. These arguments are going to be used in order to justify
that confinement is also necessary for the TC chiral symmetry breaking.
  
In Ref.\cite{cornwall} it was proposed the following scenario for QCD: a) Gluons acquire a dynamical mass, b) The theory with dynamically
massive gluons generate vortices, and c) These center vortices generate confinement. Lattice simulations are showing evidences for 
a relation between CSB and confinement, where center vortices play a fundamental role. In the $SU(2)$ case the artificial center vortices removal also implies recovery of the chiral symmetry \cite{vort1,vort2,bow1}! We also have another lattice result indicating the importance of the deep infrared region for CSB in QCD \cite{sugaa}. In this simulation the quark condensate $\left\langle {\bar{q}}q\right\rangle$ is drastically reduced ($\approx $ 40$\%$) by removing very low momentum gluons. This last result is consistent with the CSB mechanism obtained in the confinement model of Ref.\cite{cornwall3}, as shown in Ref.\cite{we}, where most of the CSB is due to gluons with momentum smaller than a few hundred MeV. Finally, continuum arguments also claim that confinement is necessary and sufficient for CSB \cite{cornwall5}. 

There is also another QCD confinement fact indicating that we need something else than the one dynamically massive gauge boson exchange
to explain the strong force. It is known for a long date that the following static potential leads to a quite successful quarkonium phenomenology
\be
V_F (r) = K_F r - \frac{4}{3} \frac{\alpha_s}{r} \,\, ,
\label{pot1}
\ee
where the first (confining) term is linear with the distance and proportional to the string tension $K_F$. The second term, that is 
of order $\alpha_s$, the strong coupling constant, describes the one gluon exchange contribution. 
The classical potential between static quark charges is related to the Fourier transform of the time-time
component of the full gluon propagator in the following way
\be
V ({\bf{r}}) = - \frac{2C_2}{\pi} \int d^3 {\bf{q}} \alpha_s ({\bf{q}}^2) \Delta_{00}({\bf{q}}) \exp^{\imath {\bf{q.r}}} \,\, ,
\label{pot12}
\ee
where the bold terms, ${\bf{q}}$ and ${\bf{r}}$, are 3-vectors. As noticed in Ref.\cite{vento} \textit{the linear confining term of the potential ($K_F r$) cannot be obtained from the gluon propagator determined in the lattice or from the gluonic SDE}, i.e. we could roughly say that the dynamically massive gluon propagator also does not lead to quark confinement as it may not lead to CSB for fermions in the fundamental representation.
The existence of a linear confining potential felt by quarks 
is supported by lattice simulations \cite{greensite}, and is a strong justificative for a confining effective propagator. 
This linear confining part of the potential must also show a cutoff at some distance. For $n_f =2$ quarks in the fundamental representation, lattice QCD data seems to indicate that the string breaks at the following critical distance \cite{bali} 
\be
r_c \approx 1.25 \,\,\, fm \,\, ,
\label{critm}
\ee
which corresponds to a critical mass (or momentum), compatible with the $m$ value necessary to generate the expected amount of CSB in the gap equation. This distance may change with the fermionic representation (because the string tension changes with the fermionic representation \cite{greensite}), but there shall always be a critical value associated to the string breaking or to the force screening. 

All the above facts were collected in order to show that a theory with dynamically massive gauge bosons, as expected for
any asymptotically free non-Abelian gauge theory, may not have enough strength to generate CSB with fermions in the fundamental
representation. Of course, for large fermionic representations, with a large value for the Casimir operator [$C_2$ in Eq.(\ref{eqn1})]
this may not be true \cite{cornwall2}. Confinement and CSB
seem to be intimately connected. The Fourier transform (Eq.(\ref{pot12})) of a dynamically massive gauge boson propagator does not 
lead to a confining potential ($\propto K_F r$), although it can explain its short distance behavior ($\propto \alpha_s / r$) \cite{vento}.
In some way confinement must also be limited to some scale as described by Eq.(\ref{critm}).
Therefore, to model CSB in QCD or TC, as we intend to do in a Schwinger-Dyson equation approach, we must introduce confinement explicitly and also consider the one-gauge dynamically massive boson exchange. The propagators that we shall use in the fermionic Schwinger-Dyson equation, when plugged into Eq.(\ref{pot12}), have to reproduce at some extent the behavior of Eq.(\ref{pot1}) and the confining contribution has to reflect the limit
shown in Eq.(\ref{critm}). These ideas that were introduced in Ref.\cite{cornwall3} and applied phenomenologically in Ref.\cite{we} in the QCD case, are going to be extended to TC theories in this work.

Cornwall introduced a confinement effect explicitly into the gap equation through the 
following {\it effective propagator}, {\it which is not at all related to the propagation of a standard quantum field} \cite{cornwall3}:
\be
D_{eff}^{\mu \nu}(k) \equiv \delta^{\mu \nu} D_{eff} (k); \,\,\,\,\,  D_{eff} (k)=\frac{8\pi K_F}{(k^2+m^2)^2}   \, ,
\label{eq01}
\ee
where $K_F$ is the string tension. In\, the\,\, $m\rightarrow 0$\,\, limit\, we\, would\, obtain\, the\, standard\, effective\, propagator $8\pi K_F\delta^{\mu\nu}/k^4$ 
that yields approximately an area law for the Wilson loop. This propagator has an Abelian gauge invariance that appears in the quark action obtained 
by integrating over quark world lines implying an area-law action \cite{cornwall3}. 
We must necessarily have a finite $m\neq 0$ value due to entropic reasons as demonstrated in Ref.\cite{cornwall3}, and its value is related to the dynamical quark mass ($m\approx M(0)$), as required by gauge invariance, originating a negative term $-K_F/m$  in the static potential in order to generate the Goldstone bosons associated to the chiral symmetry breaking. 

We can now turn to TC and write down what we may expect for the gap equation. As happens in the QCD case, the technifermion SDE can be modeled by the sum of a part containing the confining effective propagator plus another contribution with a massive one-techni-gluon exchange \cite{cornwall3,we}, which, in the Abelian techni-gluon approximation, is given by 
\br
&& M(p^2)=\frac{1}{(2\pi)^4}\int \, d^4k \, D^{tc}_{eff} (p-k) \frac{4M(k^2)}{k^2+M^2(k^2)}  + \nonumber \\
&& \hspace{-0.3cm}\frac{C_2 }{(2\pi)^4}\int  d^4k   \frac{{\bar{g}_{tc}}^2(p-k)3M(k^2)}{[(p-k)^2+m_{tg}^2(p-k)][k^2+M^2(k^2)]}  ,
\label{eq0511}
\er
where $M(p^2)=M_c(p^2)+M_{1tg} (p^2)$ is the dynamical techni-quark mass generated by the effective confining and the dressed techni-gluon 
contributions. This last equation is the basic one that we shall explore in this work. Note that the effective propagator in the
first integral of Eq.(\ref{eq0511}) leads to a confining potential ($\propto K_{tc} r$) and the massive techni-gluon exchange to the
short distance contribution ($\propto \alpha_{tc} /r$) of the static TC potential. We have just replaced the QCD quantities ($K_F$, 
${\bar{g}}^2_{QCD}$ and $m_g$) by the equivalent TC quantities ($K_{tc}$, 
${\bar{g}}^2_{tc}$ and $m_{tg}$). In the following we also assume that the string tension in the confining
propagator has also to be changed according to the fermionic representation \cite{we}, but much of our discussion will be related to
fermions in the fundamental representation.

If the TC theory contains fermions in the fundamental representation it can be shown that just the first integral on the right hand side of Eq.(\ref{eq0511}), i.e. the gap equation without the
massive technigluon exchange, is enough to generate the desirable amount of chiral symmetry breaking (with appropriate values $K_{tc}$ and $m\approx M(0)$). The asymptotic behavior of the self-energy in this case is
\[
M(p^2)|_{p^2\rightarrow\infty}\propto {1}/{p^4} \,\, ,
\]
which is a very soft behavior. The one-technigluon exchange enters only to modify the asymptotic behavior of the gap equation
as happens in the QCD case \cite{we}.

The full gap equation can be transformed into a differential equation and it is possible to verify that the solution is a linear combination of two independent solutions of the form $f(x)= b_1 f_{reg}(x) + b_2 f_{irreg}(x)$, where $b_1$ and $b_2$ are determined by the boundary conditions.
The asymptotic behavior is dominated by the one-technigluon exchange contribution, whereas \textit{the effects of the confining propagator enter only through the boundary conditions} \cite{we}. 
In Ref.\cite{we} we verified that the irregular solution
dominates when a cutoff $\Lambda \approx m$ is introduced. In a $SU(N)$ technicolor theory this ultraviolet behavior would be
equal to \cite{we}
\be
M(p^2)|_{p^2\rightarrow\infty}\propto M (\ln{p^2/M^2})^{-d}\,\, ,
\label{eqir}
\ee 
where $d=9C_2/(11N-2n_f)$ for $n_f$ flavors. This solution minimizes the vacuum energy
and has a vacuum expectation value proportional to $1/g^2$ \cite{nat}. All the above comment is just to recall how the boundary conditions
may affect the behavior of the self-energy. We shall not consider $\Lambda \approx m$ in the sequence, but we will argue that the integrals
in Eq.(\ref{eq0511}) should be performed in different momentum regions.

We now suppose that the confining propagator is limited to a specific momentum interval.
The confining propagator that we are discussing here is not the one of a fundamental field, therefore we argue that it must specify a certain
region where confinement should exist. If the string breaking happens at a certain critical distance ($r_c$),
and if the phenomenological classical potential between static quark charges is given by the Fourier transform of the time-time component of this confining propagator, the confining propagator will not reflect this breaking unless we cut the momentum up to a maximum value where the confinement region
exists, or we can understand the momentum flowing in the confining propagator as the energy that may flow between confined quarks. If this hypothesis is
correct it is natural to have the following four-fermion approximation \cite{we}:  
\br
&&M(p^2) \approx M_{4f}(p^2)= \nonumber \\
&&\frac{2}{\pi^3}\!\frac{K_R}{m_{tc}^4}\! \int \!\! d^4k \, \frac{M_{4f}(k^2) \theta (m_{tc}^2-k^2) }{k^2+M_{4f}^2(k^2)}  
+\frac{C_2 }{(2\pi)^4} \times \nonumber  \\
&&\int^\Lambda \! d^4k \frac{{\bar{g}_{tc}}^2(p-k)3M_{4f}(k^2)}{[(p-k)^2+m_{tg}^2(p-k)][k^2+M_{4f}^2(k^2)]} .
\label{eq25}
\er
In Ref.\cite{we} we verified that \textit{the critical behavior of Eq.(\ref{eq25}) and the one of Eq.(\ref{eq0511}) are basically the same in
what concerns the critical values of the ``constants" $K_{F,tc}$ and $m$}, with the massive one-gauge boson exchange barely affecting the symmetry breaking. The value of the chiral parameters, like the dynamical fermion mass and condensates, are not so much different, implying that the
approximation is quite reasonable. This is a consequence of the very strong confining force and the fact that most of the symmetry breaking
is dominated by the physics at very low momenta. 

The solution of Eq.(\ref{eq25}) has a slow decrease with the momentum \cite{we} and is typical of the gauged Nambu-Jona-Lasinio (NJL) type of models \cite{takeuchi}. The dressed one-gluon exchange has not enough strength to generate such type of four-fermion interaction \cite{we}, which
occurs only due to the large ratio between the string tension and the factor $m$ in the confining potential. Actually, we
have a simple reasoning to explain why the self-energy solution is the one corresponding to what is called irregular
behavior (or NJL type of solution).  Eq.(\ref{eq0511}) is a particular case of the following equation:
\br
M(p^2) \approx && \beta \! \int^{m^2} \!\! d^2k k^2 \, G(p,k) \frac{M(k^2) }{k^2+M^2(k^2)}  \nonumber  \\
&& \!\!\!\!\! + \alpha \int^{\Lambda^2} \! d^4k \,  \frac{{\bar{g}}^2(p-k)M(k^2)}{[(p-k)^2][k^2+M^2(k^2)]} ,
\label{eq26}
\er
where $\Lambda$ is an ultraviolet cutoff, $G(p,k)$ is an integrable function in the interval $[0,m^2]$, where the interval is understood for $p$ and $k$, and we have chosen arbitrarily $m$ as the momentum limit to where confinement is propagated. 
$M(k^2)$ is a well behaved function in the infrared with $M(0)\approx m$.
We can verify that the ultraviolet boundary condition behavior ($p^2 \rightarrow \Lambda^2$) of Eq.(\ref{eq26}) is given by
\be
M(\Lambda^2) \propto \beta \int^{m^2} \!\! d^2k k^2 \, \frac{M(k^2) }{k^2+M^2(k^2)}  \,\, ,
\ee
which is a constant and not different from a bare mass in the gap equation, leading to a hard behavior for the dynamical mass. 
Another argument in favor of limiting the confining propagator to a certain momentum region can also be abstracted from Ref.\cite{bs}
and references therein, although we do not need necessarily to interpret the quark condensate that is generated in our case as an ``inside
hadron" condensate. 

It is known that the introduction of a four-fermion interaction is responsible for harder self-energy solutions in non-Abelian gauge theories \cite{takeuchi}. In our case this four-fermion interaction is natural because of the
enormous strength of the effective confining propagator. We can give more arguments about the four-fermion approximation, which also
support the introduction of an effective confining propagator: 
Four-fermion interactions are known for a long date to describe the low energy strong interaction behavior, and it would be quite difficult to imagine that only a massive (techni)gluon propagator could lead to an effective four-fermion interaction,
because the actual interaction strength for the perturbative gap equation 
is measured by the product ``coupling$\otimes$propagator", and we know from Eq.(\ref{eq04}) 
that the $1$-(techni)gluon exchange has not enough strength to generate such effective coupling. On the other hand the confining effective 
propagator, with the usual values for the string tension, is strong enough to generate the effective gap equation (\ref{eq25})!
Apart from the (techni)gluon mass effect appearing in the $1$-(techni)\-gluon contribution, Eq.(\ref{eq25}) has been extensively 
studied in Ref. \cite{takeuchi}, and it does lead to a self-energy solution that decreases slowly with
the momentum, although the origin and the \textit{cutoffs}
are totally different. \textit{This can also be verified
comparing the $4$-fermion coupling constant ($\lambda$) of Ref.\cite{takeuchi} with our ``effective coupling" $K_R /m_{tc}^2$}, related
to the representation $R$ of the TC group.
The fermion condensate in a given representation $R$ obtained from Eq.(\ref{eq25}), as shown in Ref.\cite{we}, has the same form found by Takeuchi (Eq.(6) of Ref.\cite{takeuchi})
\be
\left\langle {\bar{q}}q\right\rangle_R (m_{tc}^2) \approx - \frac{N_R}{8\pi} \frac{m_{tc}^4}{K_R} M_R(m_{tc}^2),
\ee
corresponding to a broken-symmetry phase characterized by $K_R /m_{tc}^2 >1$ (or $\lambda >1$ in Fig.(1) of the first paper in \cite{takeuchi}), leading naturally to large anomalous dimensions produced by the confining propagator and a very hard dynamics for the self-energy.
Therefore, there is no
reason to expect a different behavior in a TC model, or any asymptotically free non-Abelian theory, as long as the theory is in
the confining phase \cite{we}.  

Summarizing our discussion we can say that the explicit introduction of confinement into the gap equation \cite{cornwall3} gives a possible solution for the problem of CSB when the gauge bosons have
a dynamically generated mass. It is necessary to generate the linear potential as well as to promote the symmetry breaking associated
to the deep infrared region, which are facts observed in lattice simulations \cite{vort1,vort2,bow1,sugaa}. The introduction of the
scale $m\approx M(0)$ into the confining propagator is necessary for entropic reasons, otherwise it would be extremely difficult
to generate the Goldstone bosons associated to the chiral symmetry breaking \cite{cornwall3}. Within the approximations discussed here
these results are valid for any non-Abelian gauge theory in the confining phase.

\section{Technicolor models with dynamically massive technigluons and confinement effects}

\par It is possible to outline a class of TC models that can be built based on the irregular solution for the self-energy, with \textit{the main advantage that QCD and TC have the same type of self-energy and participate equally in the mass generation mechanism for the known fermi\-ons} \cite{doff2}.  
Considering that QCD and TC have the so called ``irregular" self-energy \cite{we,takeuchi}, which will be parameterized as \cite{doff2,doff3}
\be 
\Sigma (p^2) \sim \mu \left[1 + b g^2 \ln\left(p^2/ \mu^2 \right) \right]^{-\gamma }  \,\,\, ,
\label{eq12}
\ee	
where\, $\mu$ \,is \,the characteristic scale of mass generation (QCD or TC),\, $\gamma= 3c/16\pi^2 b$\, and
$c = \frac{1}{2}\left[C_{2}(R_{1}) +  C_{2}(R_{2}) - C_{2}(R_{3})\right].$ 
$C_{2}(R_{i})$ are the Casimir operators for fermions in the representations  $R_{1}$ and $R_{2}$ that condense in the representation $R_{3}$,
when we compute the ordinary fermion mass ($m_f$) we obtain \cite{doff2}:
\br
&&m_f \approx g^2_{etc} \mu_{TC(QCD)} \times \nonumber \\
&& \left[1 + b_{TC(QCD)} g_{TC(QCD)}^2 \ln\left(M_{etc}^2/ \mu_{TC(QCD)}^2 \right) \right]^{-\gamma }.  
\label{mf}
\er
In the above equation $\mu_{TC(QCD)}$ is the characteristic TC(QCD) chiral symmetry breaking scale, $g^2_{etc}$ is the ETC coupling constant, $b_{TC(QCD)}$ the first $\beta$ function coefficient, $g_{TC(QCD)}^2$ is the TC(QCD) coupling constant, $M_{etc}$ the ETC boson mass, and we 
also neglected some constants. Three points are very important to be noticed: a) The fermion masses depend quite weakly on the ETC boson mass,
which may have very large values not leading to FCNC problems, b) Small fermion masses are generated when the chiral symmetry breaking is due to QCD. This is quite different from the usual models where it is assumed that
QCD has a very soft solution for the self-energy, c) The largest mass that we can generate, if $\mu_{TC}$ is of the order of a few hundred
GeV, is roughly of order of $g^2_{etc}\mu_{TC}$ and not too much different from the top quark mass \cite{doff3}.

According to the previous paragraph we can say that we may generate two different mass values for the ordinary fermions:
\be
m_f^{light} \approx g^2_{etc} \mu_{QCD} \,\,\,\,\,  , \,\,\,\,\, 
m_f^{heavy} \approx g^2_{etc} \mu_{TC} \,\, ,
\label{masi}
\ee 
where we neglected the (small) contribution of the term between brackets in Eq.(\ref{mf}). If we compute the condensates we also
can verify that we have a QCD and TC condensates with scales separated by an ${\cal{O}}(10^3)$. The light masses are of order of
the first generation fermion masses, while the heavier are of the order of the third generation masses \cite{doff2}. But how is
it possible to prevent light fermions to acquire heavy masses? This can be solved with the help of a family, or horizontal, symmetry.

In the sequence we sketch a scheme quite similar to the one proposed by Berezhiani and Gelmini {\it et al.} \cite{ref3} where their vevs of fundamental
scalars are substituted by QCD and TC condensates \cite{doff2}.
Let us suppose that we have a horizontal symmetry based on the $SU(3)_H$ group and the TC theory has technifermions in the fundamental
representation of $SU(4)_{TC}$. The technifermions form a quartet under  $SU(4)_{TC}$
and the quarks are triplets of QCD. The technicolor and color condensates will be formed at the scales
$\mu_{TC}$ and $\mu_{QCD}$ in the most attractive
channel (mac) of the products  ${\bf \bar{4}\otimes 4}$ and ${\bf \bar{3}\otimes 3}$ of each strongly interacting theory.
We assign the horizontal quantum numbers to technifermions and quarks such that these same products
can be decomposed in the following representations of  $SU(3)_H$:  ${\bf \overline{6}}$ in the case of the TC
condensate, and  ${\bf 3}$ in the case of the QCD condensate. For this it is enough that the standard
left-handed (right-handed) fermions transform as triplets (antitriplets) under $SU(3)_H$, assuming that the TC and QCD
condensates are formed in the ${\bf \overline{6}}$ and in the ${\bf 3}$ of  the $SU(3)_H$ group. This is consistent
with the mac hypothesis although a complete analysis of this problem is out
of the scope of this work.
The above choice for the condensation channels is crucial for our model, because the TC condensate in
the representation  ${\bf \overline{6}}$  (of $SU(3)_H$) will interact only with the third fermionic generation while the  ${\bf 3}$
(the QCD condensate) will interact
only with the first generation. In this way we can generate the coefficients $C$ and $A$ respectively of a Fritzsch type matrix \cite{fritzsch},
because when we add these condensates (vevs) and write them as a $3 \times 3$ matrix  we will end up ({\sl at leading order}) with
 \br
 M_f =\left(\begin{array}{ccc} 0 & A & 0\\ A^* & 0 & 0 \\
0 & 0 & C
\end{array}\right).
\label{e12} \er
The points that must still be discussed are how we generate the intermediate masses and why the contribution of the term between brackets in Eq.(\ref{mf}) is indeed small and can be neglected.

In the scenario that we shall consider the ETC group can connect all fermions and contain the TC and QCD interactions. Actually the ETC role can
be played by a grand unified theory (GUT), which has exactly these characteristics. This is possible because the fermion mass barely depend
on the ETC or GUT gauge boson masses, as can be verified from Eq.(\ref{mf}). These ETC or GUT bosons can intermediate neutral flavor changing interactions,
however they can be very heavy in order to be consistent with all experimental constraints on FCNC interactions \cite{doff2}. 
We can build a TC model based on a GUT such that
\be
G_{gut}\supset G_{SM}\otimes SU(N)_{TC}\otimes G_H \,\, ,
\label{gu}
\ee
where $G_{SM}$ is the Standard Model group, $SU(N)_{TC}$ is the TC group and $G_H$ corresponds to a horizontal symmetry, which is
not necessarily a local one, but with a characteristic scale of the order of the GUT scale, and, for simplicity, couplings are assumed 
to be of the same order (i.e. $g^2_H\approx g^2_{gut}$). TC should condensate at TeV scale. All groups are embedded into the GUT, therefore we may have all kind of neutral flavor changing interactions but at
the GUT scale, since this theory will play the role of the ETC theory \cite{sannino}.

In the example that we discussed before, where
$G_H \equiv SU(3)_H$ with technifermions condensing in the ${\bf{\bar{6}}}$ and quarks condensing in the ${\bf{3}}$ representations of the
horizontal group, we can obtain the following mass matrix \cite{doff2}
\br
 M_f =\left(\begin{array}{ccc} 0 & A & 0\\ A^* & 0 & B \\
0 & B^* & C
\end{array}\right),
\label{e1} 
\er
where $A\propto g^2_{gut} \mu_{QCD}$ $\approx {\cal{O}}$(MeV) and 
$C\propto g^2_{gut} \mu_{TC}\approx {\cal{O}}$(GeV). The $B$ term has an intermediate value naturally generated by the effective potential of the
\textit{composite} ${\bf{\bar{6}}}$ and ${\bf{3}}$ Higgs system as shown in \cite{doff2}. Notice that we can only obtain a mass as heavy
as the top quark one in TC models with the use of Eq.(\ref{eq12}) \cite{doff3}. 

To show that a mass matrix like the one of Eq.(\ref{e1}) is a feasible one, we can use the technique of effective potential for
composite operators as discussed in Ref.\cite{doff}, and verify that QCD and TC lead to a two composite Higgs boson system indicated respectively
by $\eta$ and $\phi$, with, due to the horizontal symmetry, the following vacuum expectation values (vevs):
\br
\left\langle \eta \right\rangle \approx \left(\begin{array}{c} 0 \\ 0 \\ v_\eta \end{array}\right) , \,\,\,\,\,\,
\left\langle \phi \right\rangle \approx \left(\begin{array}{ccc} 0 & 0 & 0 \\ 0 & 0 & 0\\ 0 & 0 & v_\phi \end{array}\right) ,
\er
where the first vev will be of the order of 250 MeV and the second one of order 250 GeV. The intermediate term in Eq.(\ref{e1}) will
be originated by mixed terms in the effective potential of our composite system \cite{doff2}. These terms will come out
naturally from one-loop standard model interactions connecting $\eta$ and $\phi$ (or quarks and techniquarks scalar composites),
being of the following type
\be
V_2 (\eta, \phi) = \epsilon \eta^\dagger \eta \phi^\dagger \phi + \delta \eta^\dagger \phi \eta \phi^\dagger + ...
\ee
The details of how this effective potential contribution originates in such type of models were worked out in Ref.\cite{doff2}. 

Let us summarize why we consider this a quite general type of model. First, due to the fact that confinement is responsible
for a dynamical mass typical of a NJL gauged model, or of the irregular type, we end up with two scales of ordinary fermion masses: QCD and TC.
Secondly, due to the form of the self-energy the fermion masses barely depend on the ETC gauge boson masses, and these can be quite
heavy and do not generate FCNC problems. Finally, to generate reasonable fermion mass matrix we only need a horizontal or family symmetry.
There are possibly many theories based on different groups that may fit into this scheme, and we do not need to appeal to technifermions
that belong to higher TC representations.

\section{Conclusions}

\par We initiated our work calling attention to the fact that in asymptotically free non-Abelian gauge theories the gauge bosons may
acquire naturally dynamical masses. This fact has been already checked for QCD through lattice simulations and Schwinger-Dyson
equations. We expect that the same phenomenon occurs in TC theories if the theory has not too many fermions in order to spoil
the gauge boson mass generation mechanism. Chiral symmetry breaking in TC model with dynamically generated technigluon masses
was discussed here, as far as we know, for the first time. We argue, based on QCD results, that with dynamically massive technigluons
it may be quite difficult to promote chiral symmetry breaking in TC theories, particularly if technifermions are in the
fundamental representation of the TC group.

Based on lattice results and continuum arguments we followed Cornwall's idea that confinement is necessary and sufficient
for chiral symmetry breaking \cite{cornwall3,cornwall4}. Confinement has to be introduced explicitly into the fermionic Schwinger-Dyson equation. This is
performed with the introduction of an effective confining propagator, in a way quite similar to the proposition of the phenomenological static potential
of Eq.(\ref{pot1}), which is quite successful in describing the quarkonium spectra. The new gap equation, in the QCD case, was extensively
discussed in Refs.\cite{cornwall3,cornwall4,we}, and we just rewrite it in the TC case.

The TC gap equation containing a confining propagator and dynamically massive technigluons is discussed following the
steps already pointed out in Ref.\cite{we}. The main point is the introduction of an infrared cutoff in the confining part
of the gap equation. The result is that the confining propagator is responsible to generate a contribution typical of a bare mass
in the fermionic Schwinger-Dyson equation, what leads to a very ``hard" self-energy, or a self-energy of the irregular type.
One important fact is that the full gap equation is well approximated by a four-fermion interaction, and its critical
behavior is not different from the one of the full equation, as shown in Ref.\cite{we}. Moreover, the numerical values for the chiral parameters
obtained with the four-fermion approximation do not differ from the ones of the full equation \cite{we}. In TC we also
should expect the same four-fermion interaction as happens in QCD; they appear here in the same way that they appear in
walking TC theories \cite{takeuchi}, although their origin in our case is totally different and based on the confinement
effect.

As a consequence of confinement and dynamical gau\-ge boson mass generation, leading to a very particular fermionic self-energy,
we see that the chiral symmetry breaking of both theories, QCD and TC, participate in the generation of ordinary fermion masses.   
These fermion masses barely depend on the ETC mass scale. This allow us to build a quite general type of TC model, where the
ETC interaction can be naturally substituted by a GUT interaction, at the cost of introducing a horizontal or family symmetry
to prevent light fermions of acquiring masses directly from the TC condensates. The new gauge boson interactions (horizontal or GUT)  
appear at a very high energy scale and we do not expect FCNC at undesirable levels in this type of model. 

The fact that most of the first fermionic family masses 
are originated from the QCD chiral symmetry breaking and the third fermionic family masses comes from the TC chiral breaking is a
novelty. This is a direct consequence from the possibility that confinement, in non-Abelian gauge theories with fermions in the fundamental representation, induces an effective four-fermion interaction simulating a bare mass, where the self-energy
decreases very slowly with the momentum and may be a solution for the phenomenological TC problems.

\ack
This research was partially supported by the Conselho Nacional de Desenvolvimento
Cient\'{\i}fico e Tecnol\'ogico (CNPq) (AD and AAN) and by Funda\c c\~ao de Amparo a
Pesquisa do Estado de S\~ao Paulo (FAPESP) (FAM).

\end{document}